\documentclass[
  journal=largetwo,
  manuscript=article-type,
  year=2020,
  volume=37,
]{cup-journal}

\usepackage{amsmath}
\usepackage[nopatch]{microtype}
\usepackage{booktabs}
\usepackage{hyperref} 
\usepackage{amssymb}

\hypersetup{colorlinks,citecolor=blue,linkcolor=blue,urlcolor=blue}

\title{A Near-Field Treatment of Aperture Synthesis Techniques using the Murchison Widefield Array}

\author{S. Prabu$^1$}
\affiliation{International Centre for Radio Astronomy Research, Curtin University, Bentley, WA 6102, Australia}

\email[S. Prabu]{steveraj.prabu@curtin.edu.au}

\author{S.J. Tingay$^1$}

\author{A. Williams$^1$}


\begin{document}

\begin{abstract}
Typical radio interferometer observations are performed assuming the source of radiation to be in the far-field of the instrument, resulting in a two-dimensional Fourier relationship between the observed visibilities in the aperture plane and the sky brightness distribution (over a small field of view). When near-field objects are present in an observation, the standard approach applies far-field delays during correlation, resulting in loss of signal coherence for the signal from the near-field object. In this paper, we demonstrate near-field aperture synthesis techniques using a Murchison Widefield Array observation of the International Space Station (ISS), as it appears as a bright near-field object. We perform visibility phase corrections to restore coherence across the array for the near-field object (however not restoring coherence losses due to time and frequency averaging at the correlator). We illustrate the impact of the near-field corrections in the aperture plane and the sky plane. The aperture plane curves to match the curvature of the near-field wavefront, and in the sky plane near-field corrections manifest as fringe rotations at different rates as we bring the focal point of the array from infinity to the desired near-field distance. We also demonstrate the inverse scenario of inferring the line-of-sight range of the ISS by inverting the apparent curvature of the wavefront seen by the aperture. We conclude the paper by briefly discussing the limitations of the methods developed and the near-field science cases where our approach can be exploited.
\end{abstract}

\section{Introduction}
Using three basic assumptions, conventional aperture synthesis theory derives a 2D Fourier relationship \citep{thompson2017interferometry, marr2015fundamentals} between the visibilities sampled by an interferometer in the aperture plane and the sky brightness distribution.  The three assumptions are: that a narrow bandwidth is used; that the object being observed is in the far-field; and that a narrow Field of View (FOV) is being imaged. Techniques such as Multi-Frequency Synthesis \citep{sault1994multi, conway1990multi, rau2011multi} and W-Stacking \citep{offringa-wsclean-2014, offringa-wsclean-2017} have been developed to to overcome the bandwidth and FOV limits, in this paper we develop techniques/tools to observe objects in the near-field of the instrument, building upon previous work. We demonstrate near-field imaging techniques using the Murchison Widefield Array (MWA) \citep{Tingay2013TheFrequencies, 2018PASA3533W}, whose long ($6$\,km) baselines see objects in Low Earth Orbit (LEO) in the near-field (for the frequency range that the MWA operates in).\\

The MWA is a radio interferometer built as a precursor to the low-frequency component of the Square Kilometre Array (SKA) and is located in the radio-quiet region of the Murchison Shire in Western Australia, at \textit{Inyarrimunha Ilgari Bundara}, the CSIRO Murchison Radio-astronomy Observatory. The MWA is capable of observing between $80-300$\,MHz with an instantaneous bandwidth of $30.72$,\,MHz (each element in the interferometer is an aperture array composed of dual-polarised bow-tie antennas arranged in a 4x4 format, a so-called ``tile''). \\

\subsection{Previous work}
Aperture synthesis studies in the near field have been performed by many groups with varying objectives. While often signals in the near-field are considered a source of interference in astronomical observations \citep{wang2021satellite, tingay2020survey}, \cite{2023arXiv230104188F} has recently developed techniques to use satellite signals to perform calibration of the instrument. Spacecraft enthusiasts among the VLBI community 
have successfully been able to track the near-field transmissions from satellites within the solar system \citep{duev2012spacecraft, 2007IEEEP..95.2193L}. In the `ultra'\footnote{distances that are comparable to the size of the aperture array.} near-field, RFIs have also been localized to nearby transmission lines\footnote{\url{https://www.ursi.org/proceedings/procGA05/pdf/JE.5(01141).pdf}}, electric cars being charged\footnote{\url{https://www.astron.nl/dailyimage/main.php?date=20210802}}, and new LED lamps being installed in nearby farmhouses\footnote{\url{https://www.astron.nl/dailyimage/main.php?date=20200512}}. \\

In this paper, we expand on our preliminary near-field work discussed in \cite{2022AdSpR..70..812P}, and develop more versatile tools that are capable of being used for a wide range of science cases. As our previous work \citep{Tingay2013OnFeasibility,Zhang2018LimitsMWA,prabu_dev,prabu_survey} has demonstrated the MWA to be capable of detecting FM reflections from LEO satellites, we use a single $2$\, min observation of the International Space Station (ISS) as the near-field target of choice in this work. Using our developed near-field correction tools, we then infer the range of the ISS across multiple time steps by inverting the focal distance that provided the maximum signal to noise on the source. \\


This paper is structured as follows. We describe the data used and our methods in Section \ref{sec:obsandmethods}, and our results in Section \ref{sec:results}. The discussion and conclusions are in Section \ref{sec:discussion} and \ref{sec:conclusion}, respectively. \\

\begin{figure*}
\begin{center}
\includegraphics[width=0.7\linewidth,keepaspectratio]{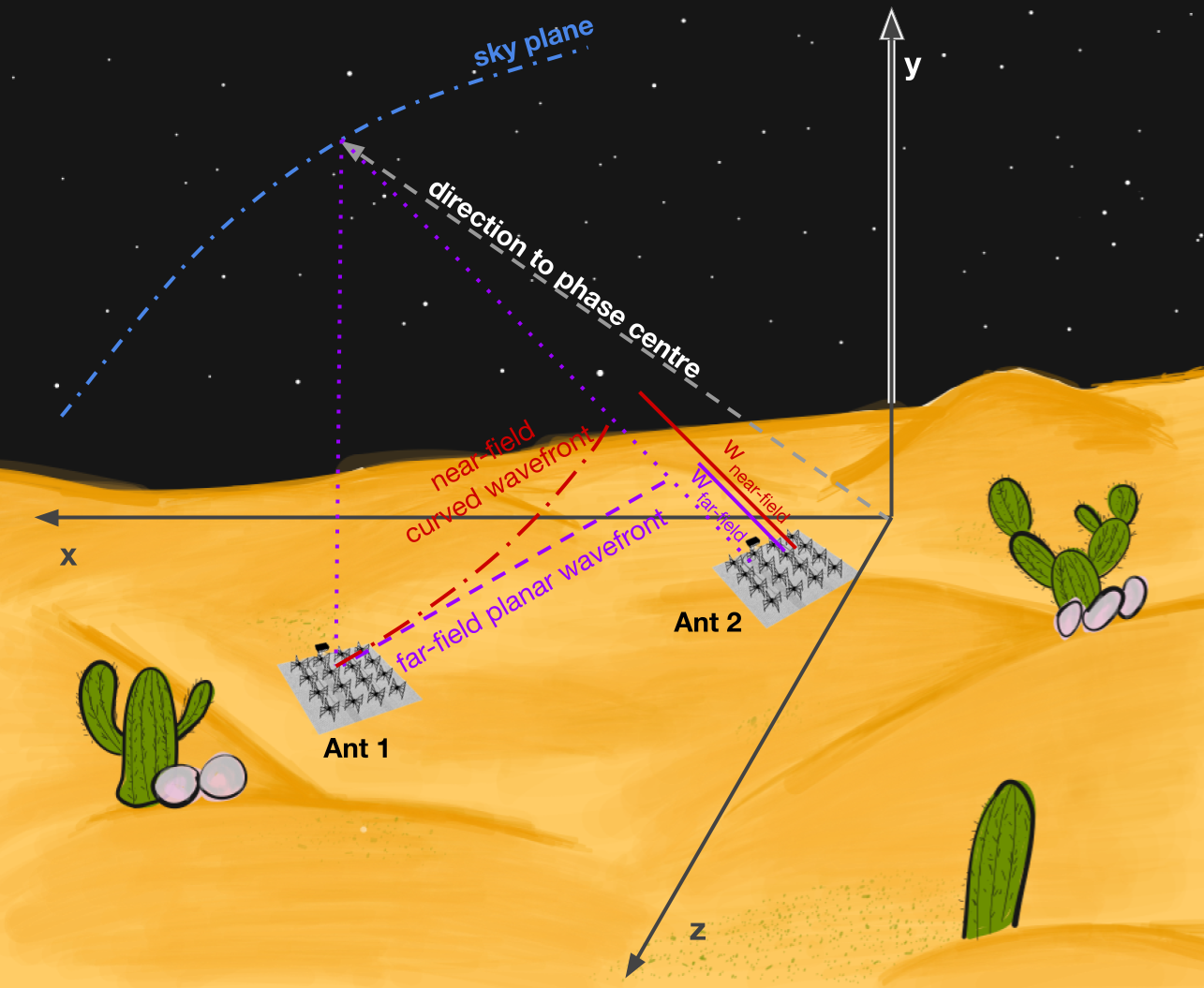}
\caption{The Topocentric Cartesian Coordinate (TCC) system used in this work to calculate near-field corrections.}
\label{FigDiagram}
\end{center}
\end{figure*}

\begin{figure}
\begin{center}
\includegraphics[width=\linewidth,keepaspectratio]{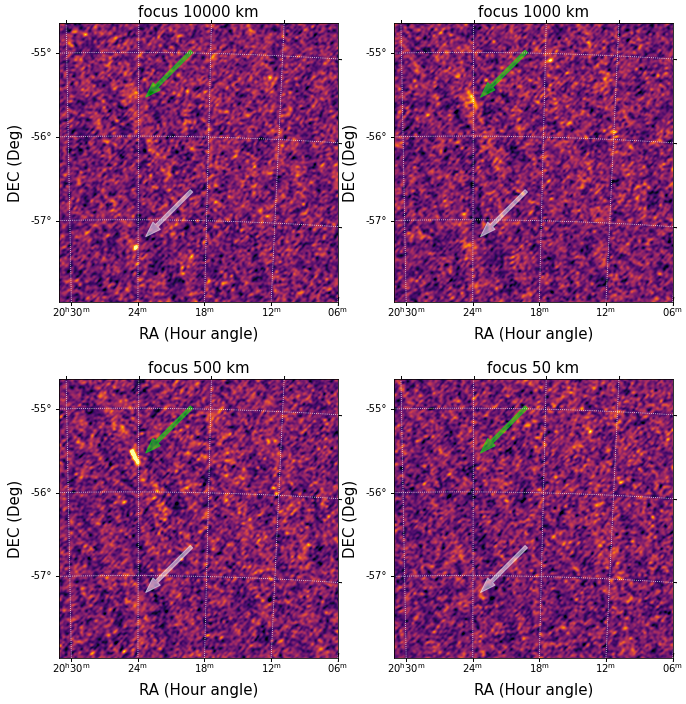}
\caption{MWA images for four different focal distances. In all four panels, the green arrows show the location of the ISS signal and the white arrow shows the location of a background astronomical point source. We note that at large focal distances, the objects in the near-field (e.g ISS) appear de-correlated and at much closer focal distances the background far-field sources appear de-correlated. An animation of this Figure is available at \url{https://www.youtube.com/watch?v=sqieJJYJCAo}. }
\label{FigNearFieldDemo}
\end{center}
\end{figure}

\section{Observations and Methods}
\label{sec:obsandmethods}
\subsection{Data Pre-Processing and Calibration}
We use a single $2$\,min Phase 3 MWA observation (obs ID 1333747192, and henceforth referred to as the target observation) of the ISS in the FM band. We downloaded the data from the All-Sky Virtual Observatory\footnote{\url{https://asvo.mwatelescope.org/}} (ASVO), with $0.25$\,s time-averaging and $40$\,kHz frequency averaging. The backend of ASVO uses {\tt Birli}\footnote{\url{https://github.com/MWATelescope/Birli}}  to convert MWA correlator output files to the {\tt CASA} defined measurement set format \citep{2007ASPC..376..127M}.\\

We calibrate the instrument by using an observation (obs ID 1333747784) of the radio galaxy Hercules-A, also downloaded from ASVO using the same parameters as the target observation. We  perform RFI flagging on the calibration observation using the {\tt AOFLAGGER} tool \citep{2015PASA...32....8O}. Using the {\tt calibrate} tool \citep{2016MNRAS.458.1057O} we perform a preliminary round of calibration using the source model\footnote{\url{https://github.com/StevePrabu/MWA-ORBITAL/blob/master/models/model-HerA-27comp_withalpha.txt}} of the Hercules-A radio galaxy. As Hercules-A is the brightest source in the FOV of the observation, we obtain a reasonably good initial amplitude and phase calibration solution. Using {\tt WSClean}\citep{offringa-wsclean-2014, offringa-wsclean-2017} and the preliminary calibration solution, we perform a round of self-calibration to obtain better calibration solutions. The final set of calibration solutions are then transferred to the target observation. \\

\begin{figure*}
\begin{center}
\includegraphics[width=0.9\linewidth]{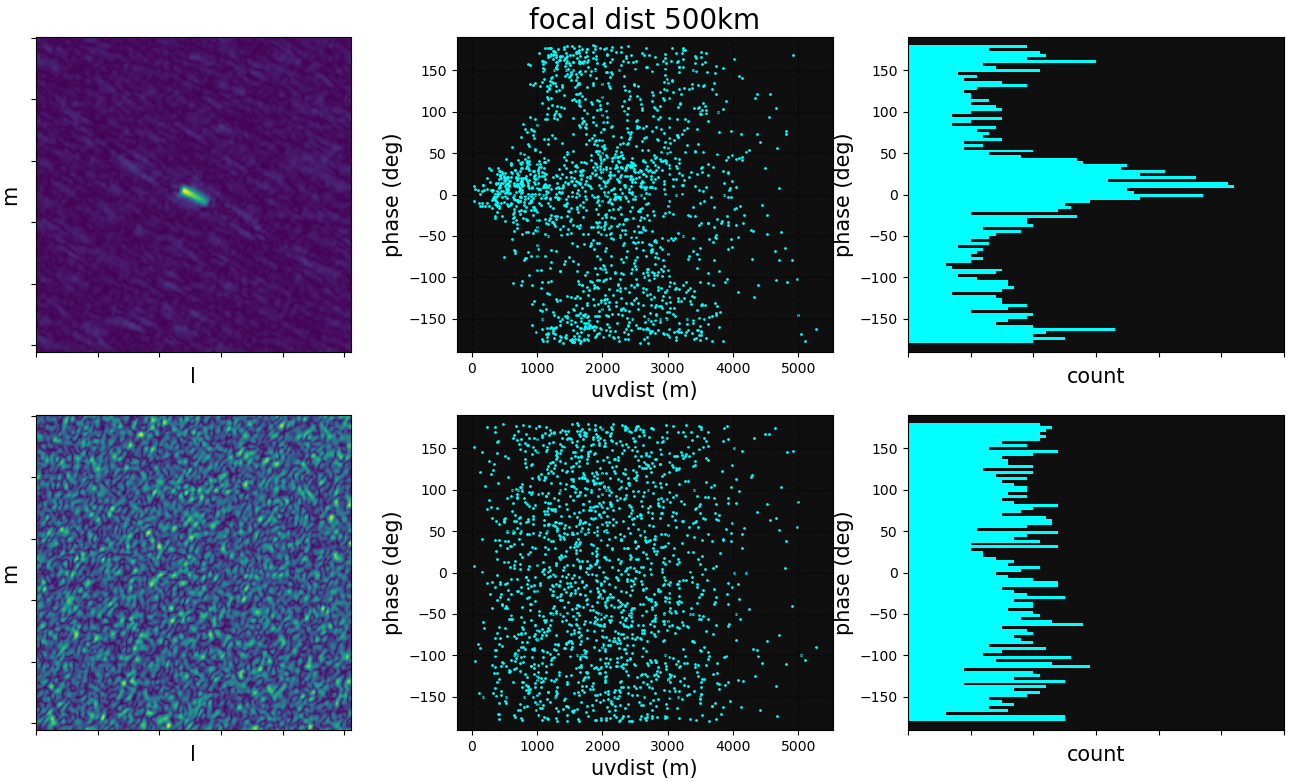}
\caption{The top three panels show the reconstructed image, visibility phases plotted against baseline length, and histogram distribution of the visibility phases for the frequency differenced visibilities obtained from differencing a channel with ISS FM signal from an adjacent channel with no FM signal. The bottom three panels show the same but for frequency difference visibilities obtained for two channels, neither of which had an ISS reflection signal. An animation of the figure for a wide range of focal distances can be obtained at \url{https://www.youtube.com/watch?v=hqPI-iFX6bY}}
\label{FigNullTest}
\end{center}
\end{figure*}

\subsection{Near-field Correction using LEOLens}
The radiation wavefronts originating from near-field objects appear curved when viewed using the long baselines of the MWA. During correlation most interferometers, including the MWA, assume the sources to be in the far field of the instrument, thus resulting in a loss of coherence in the near-field signal due to de-correlation. However, we can recover near-field phase coherence by re-arranging the fringes (phase rotation of visibilities) projected by the interferometer in the plane of the sky. We apply this near-field correction to the target visibilities using a {\tt python casacore}\footnote{\url{https://github.com/casacore/python-casacore}} tool we call {\tt LEOLens}\footnote{\url{https://github.com/StevePrabu/LEOLens}}, explained in the following paragraphs. \\ 

In Figure \ref{FigDiagram}, we show a Topocentric Cartesian Coordinate (TCC) system centered at the geometrical centre of an interferometer, such that $\hat{x}$ points East, $\hat{y}$ points towards the zenith, and $\hat{z}$ points North. Having obtained the coordinates of every MWA tile in this coordinate system, we next obtain the coordinates of the near-field object. The TCC coordinates of the near-field object ($f_{x}$, $f_{y}$, $f_{z}$) can be obtained from its azimuth angle ($\theta$), elevation angle ($\phi$), and range ($f_{dist}$) from the origin, \\

\begin{equation}
    \begin{array}{l}
        f_{x} = f_{dist} \times cos(\phi) \times sin(\theta) ,\\
        f_{y} = f_{dist} \times sin(\phi) ,\\
        f_{z} = f_{dist} \times cos(\phi) \times cos(\theta) .
    \end{array}
    \label{E1}
\end{equation}

For every baseline between antennas $A_{i}$ ($X_{i}, Y_{i}, Z_{i}$) and $A_{j}$ ($X_{j}, Y_{j}, Z_{j}$), the delay (or the w-term) for the baseline can be obtained using Equation 2. \\

\begin{equation}
    \begin{array}{l}
        r_{i} = \sqrt{(f_{x} - X_{i})^2 + (f_{y} - Y_{i})^2 + (f_{z} - Z_{i})^2} ,\\
        r_{j} = \sqrt{(f_{x} - X_{j})^2 + (f_{y} - Y_{j})^2 + (f_{z} - Z_{j})^2} ,\\
        w_{near-field,i,j} = r_{j} - r_{i} .
    \end{array}
    \label{E2}
\end{equation}

As the voltage streams from both antennas are already correlated using far-field delays ($w_{far-field,i,j}$) by the correlator, {\tt LEOLens} updates every visibility measurement using the near-field delay. \\

The capacity to define a delay (or w-term) for a baseline comes from having set a phase reference in the sky plane. Hence, any changes in the w-term (such as due to the near-field corrections done here) will rotate the baseline phase. This phase correction is applied by {\tt LEOLens} to the visibility using Equation 3. \\

 \begin{equation}
         \Delta \phi_{i,j} = \exp^{i 2\pi \frac{\Delta w_{i,j}}{\lambda}},
    \label{E2}
\end{equation}

where $\phi$ is the interferometer phase, $\lambda$ is the wavelength of radiation, and $\Delta w_{i,j} = w_{near-field,i,j} - w_{far-field,i,j}$. $\Delta \phi_{i,j}$ is the phase difference applied to the visibility phase on baseline formed by antennas $A_{i}$ and $A_{j}$.  We  illustrate the impact of near-field corrections in the aperture plane and the sky plane in \ref{sec:uvwvslm}.

\section{Results}
\label{sec:results}
Having described our near-field correction method, we present our results in three steps. In Section \ref{sec:nearfieldimages}, we show the MWA near-field images made for a wide range of focal distances, followed by a null test in Section \ref{sec:thenearfieldnulltest}. Having developed confidence in the method using the null test, we then proceed to demonstrate results from our range estimation method in Section \ref{sec:nearfieldeventrangeinference}. \\

\subsection{Near-Field Images}
\label{sec:nearfieldimages}
Using a single time-step during which the ISS was detected through FM reflection, we show near-field images at varying focal distances (Figure \ref{FigNearFieldDemo}). We focus the array over a wide range of distances ($10,000$\,km, $1000$\,km, $500$\,km, and $50$\,km) using {\tt LEOLens} and then create images using {\tt WSClean}. A distance of $10,000$\,km is in the far-field of the instrument, and hence the image is not noticeably different from the image produced without any near-field corrections. As we bring the focus of the array to much closer distances, at about $500$\,km we see a streak-like signal from the ISS, as coherence is obtained.  The ISS signal is again de-correlated as we bring the focal distance to $50$\,km. \\

In the 10,000\,km image of Figure \ref{FigNearFieldDemo}, we also see a point source (background radio galaxy) whose location in the sky is shown using the white arrow. We note that the point source is de-correlated as we bring the focal distance to smaller distances, in accordance with our expectations. Due to the source being un-resolved, all baselines respond equally to the point source across the aperture plane. As the Phase 3 MWA extended array has predominantly long baselines which undergo significant delay correction (or rotation of fringes in the sky) as we change the focal distance, the point source is de-correlated. Conversely, the overall phase structure of any extended source in the observation is expected to remain preserved for a wider range of focal distances, as the structure of an extended source is sampled by the shorter baselines that do not undergo significant delay corrections\footnote{an animation of near-field corrections applied to an extended source can be found here \url{https://www.youtube.com/watch?v=sqieJJYJCAo}. As the observation used in the animation is an MWA Phase 1 observation, and was processed differently to the other observations used here, we do not put it in the main body of the manuscript. }.

\subsection{The Near-Field Null Test}
\label{sec:thenearfieldnulltest}
We build confidence in our near-field techniques by performing a null test described below. We select a fine-frequency channel ($\nu_{1}$) containing the ISS FM reflection signal and difference the visibilities with an adjacent fine-frequency channel ($\nu_{2}$) that did not have any FM reflection. Doing so isolates the signal of interest (ISS FM reflection) from the background astronomical sources and we use it to perform the null test. Due to the close proximity of the two selected channels, the instrument's response to the background sky ($S_{sky}$ below) can be considered to be identical, the difference subtracting the sky's contribution to the measured visibilities.   Also, due to the closeness of the two channels, the instrument's response is not noticeably chromatic and does not leave behind artifacts while differencing. Mathematically the frequency differenced visibilities can be represented as follows \\

\begin{equation}
    \begin{array}{l}
        V(\nu_{1}) = \epsilon(\nu_{1}) + S_{ISS} + S_{sky}(\nu_{1}) ,\\
        V(\nu_{2}) = \epsilon(\nu_{2}) + S_{sky}(\nu_{2}) ,\\
        \Delta V = V(\nu_{1}) - V(\nu_{2}) ,\\
        \Delta V \approx \epsilon_{1,2} + S_{ISS} \quad [\because S_{sky}(\nu_{1}) \approx S_{sky}(\nu_{2})] .\\
    \end{array}
    \label{E3}
\end{equation}

In the above equations, due to the noise [$\epsilon(\nu_{1})$ and $\epsilon(\nu_{2}$)] in the two channels being un-correlated, the ISS signal ($S_{ISS}$) should be detected above a Gaussian noise distribution. We focus the differenced visibilities to $500$\,km (approx. range of the ISS) and show the corresponding image and phase-distrbution of the visibilities in the top three panels of Figure \ref{FigNullTest}. Due to the ISS signal being at the phase centre of the image, the phases of the differenced visibilities cluster near $0^{\circ}$deg and $180^{\circ}$deg, as expected. \\

In order to test the reliability of our near-field correction technique, we create a different set of frequency differenced visibilities, but this time neither of the two channels show significant ISS signal, and we focus the differenced visibilities to $500$\,km as before. As neither of the two channels have ISS signal in them, we expect the differenced visibilities to show noise-like properties. The image and the phase distribution of this new set of differenced visibilities are shown in the bottom three panels of Figure \ref{FigNullTest}. From Figure \ref{FigNullTest} we see that the phases of the visibilities are randomly distributed (as would be expected for noise) and show no coherent signals in the reconstructed image, thus building confidence in the near-field techniques/software developed in this work. \\

\begin{figure*}
\begin{center}
\includegraphics[width=0.8\linewidth]{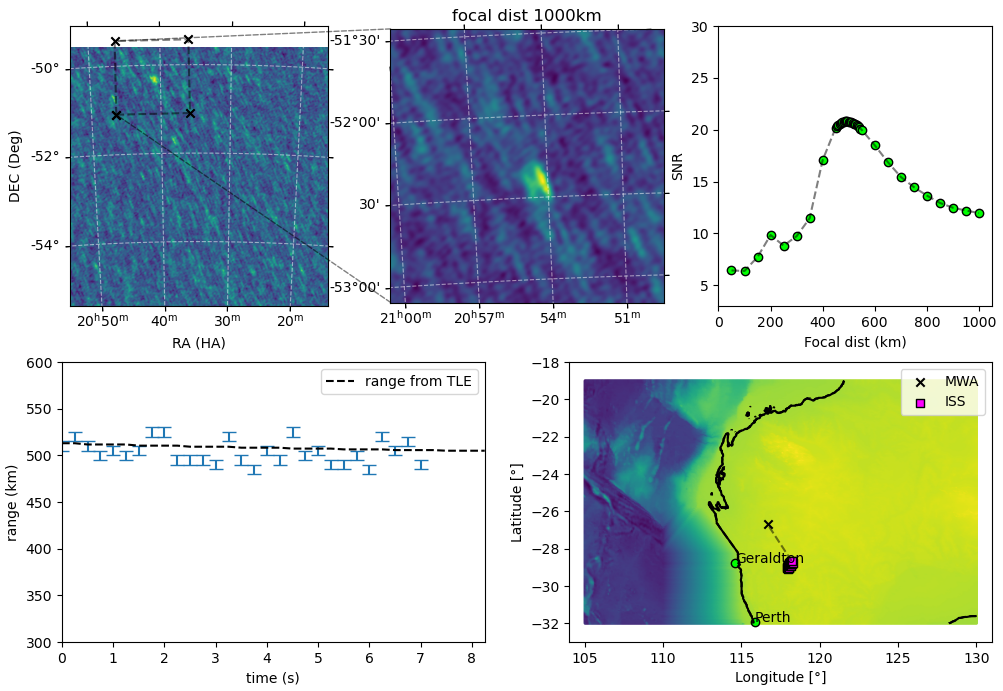}
\caption{The top-left panel shows the image with visibilities focused in the far field for one of the time steps considered. The image is also phase-centred at the default pointing centre of the observation. The insert panel (top-middle) shows the phase-tracked image of the ISS. In the top-right panel, we show the SNR of the ISS signal when focusing the array to a wide range of focal distances for a single timestep. Having performed this across multiple timesteps, we plot the estimated line of sight range in the bottom-left panel. Close to the actual range of the ISS, we attempt focussing at every 5km intervals, and hence we use $5$\,km as the error in the bottom-left plot. We also show the  Using the estimated azimuth, elevation, and range of the ISS, we are able to track its trajectory in 3D as shown in the bottom-right panel. An animation of the Figure can be obtained from \url{https://www.youtube.com/watch?v=99vksNf1viA}}
\label{FigAutoFocus}
\end{center}
\end{figure*}

\subsection{Near-Field Object Range Inference}
\label{sec:nearfieldeventrangeinference}
In Section \ref{sec:nearfieldimages}, we demonstrated being able to focus on a near-field object using prior knowledge of its distance from the geometric centre of the MWA. In this section, we demonstrate the inverse problem  of inferring the line of sight range to the object. We focus the array at a wide range of trial focal distances.  The distance that provides the maximum coherence (measured as signal-to-noise ratio (SNR) in the images) on the source is assumed to be the range to the object. \\

We demonstrate our ranging method using the frequency differenced visibilities described in Section \ref{sec:thenearfieldnulltest}. For every time-step that the ISS was detected, we change the phase-centre of the visibilities to have the ISS signal at the centre of the image. We then focus the differenced visibilities to a wide range of trial focal distances and plot the SNR of the ISS signal in the top-right panel of Figure \ref{FigAutoFocus}. As we obtain maximum coherence on the ISS signal when the assumed focal distance matches the true distance, we use the peak of the SNR vs focal distance curve as a proxy for the range measurement. In the bottom-left panel of Figure \ref{FigAutoFocus}, we plot our range measurements for the ISS across multiple time-steps and compare it to the range predicted from the Two Line Element (TLE) data for the ISS published by \url{space-track.org} during the epoch of the observation. As we know the azimuth, elevation, and range of the ISS from our observations, we are able to track the ISS trajectory in 3D space with respect to the MWA, shown in the bottom-right panel of Figure \ref{FigAutoFocus}. \\

\subsubsection{Modelling from First Principles}

The top right panel of Figure \ref{FigAutoFocus} shows that the signal recovered as a function of the assumed focal distance is quite distinctive.  Thus, while identifying the peak signal to noise of this function provides an estimate of the optimal focal distance, it would be better to model the entire function and use all of the measurement data.  We briefly consider the plausibility of forming such a model in this section.

A model to describe the data shown in Figure \ref{FigAutoFocus} needs to take into account the interferometric response of the MWA, as a function of assumed focal distance to an object.  We have tested a model that utilises the known relative positions of the MWA tiles, and calculates the interferometric response of each baseline (pair of tiles) in the array for an object at an arbitrary location with respect to those tiles, for different values of the focal distance.  In this case, errors in the delay and delay rate, relative to the true values, will cause a loss of coherence in an observation that is averaged in time and frequency.  These errors vary as a function of the length and orientation of the baseline, relative to the direction of the object.

As such, across a given time-step, the model requires three parameters to describe the starting position of the object in the array's frame of reference as well as three parameters to describe it's finishing position.  When described in spherical coordinates, the parameters can be expressed as a range, azimuth, and elevation.  The six parameter model can be fit to observational data, such as in Figure \ref{FigAutoFocus}.  In practise, the parameter space for the fit has many local minima and least squares methods struggle to approach the global minimum.  Thus, we have attempted a brute force grid search of the parameter space, which is computationally expensive.  For example, implemented in python, a single trial for a single grid point takes $\sim$10 seconds on a single CPU.  The resolution of the grid needs to be high, as the array will be coherent when all differential delays are correct to within a fraction of a wavelength ($\sim$3 m), and the grid range also needs to be large, as the inherent accuracy of our starting point from a TLE is $\sim$1 km.  Thus, that translates to $\sim$10$^{18}$ trials, or $\sim$10$^{19}$ seconds, or 100 billion years on a single CPU, which is clearly not feasible.  Optimised code would assist to reduce the compute time.

We have implemented trials using a far coarser grid as a test to see if promising regions of parameter space can be identified for further investigation, which are being run on large high performance computing clusters, and produce output that closely resembles the behaviour of the observational data.  Further refinements are a work in process.

\section{Discussion}
\label{sec:discussion}

\subsection{Visibility amplitudes}
\label{sec:amplitudefading}
While we have successfully demonstrated the recovery of phase coherence in the near-field, we have not made any comments about the amplitude of the visibilities. For astronomical sources, the amplitude of the electric field seen by both antennas of a baseline can be assumed to be identical due to the large distance to the radiating source. However, for near-field objects, the antennas see different amplitudes due to different path lengths between the object and the antennas, and is derived below. \\

Consider again a baseline between two antennas ($A_{i}$ and $A_{j}$) observing an isotropically radiating source of luminosity $L$ (J/s) at distances $r_{1}$ and $r_{2}$ from the antennas, respectively. If the effective collecting area of the two antennas are $A(l_{i}, m_{i})$ and $A(l_{j}, m_{j})$, where $l$ and $m$ are orthogonal direction cosines of the near-field source with respect to the antenna, the powers measured by the antennas are, \\

\begin{equation}
    \begin{array}{l}
        P_{i} = L \times \frac{1}{4\pi r_{i}^2} \times A(l_{i}, m_{i})\\
        P_{j} = L \times \frac{1}{4\pi r_{j}^2} \times A(l_{j}, m_{j})
    \end{array}
    \label{E4}
\end{equation}

If the radiating source is observed in a direction that is not orthogonal to the baseline, there is a propagation path length difference ($\Delta r$) between the source and the two antennas. However, due to geometrical reasons, the path length cannot exceed\footnote{not true for sources very close to the instrument with curved wave-fronts but is a good first-order guide for the magnitude of $\Delta r$.} the actual physical distance ($b$) between the two antennas (for example, the MWA's longest baseline is $6$\,km long, and $\Delta r$ for the baseline $<= 6$\,km). The two different propagation path lengths are defined as follows, \\

\begin{equation}
    \begin{array}{l}
        r_{i} = r \\
        r_{j} = r_{i,j} + \Delta r_{i,j} \quad (\mbox{where} \; \Delta r_{i,j} <= b)
    \end{array}
    \label{E5}
\end{equation}

Using Equations \ref{E4} and \ref{E5}, the ratio of powers measured by the two antennas is given by, \\

\begin{equation}
    \begin{array}{l}
        P_{ratio} = P_{i}/P_{j} \\
         P_{ratio} = \frac{A(l_{i}, m_{i})}{A(l_{j}, m_{j})} \times \frac{r_{j}^2}{r_{i}^2}\\
         P_{ratio} = \frac{A(l_{i}, m_{i})}{A(l_{j}, m_{j})}  \times \left[\frac{r + \Delta r}{r}\right]^2
        
    \end{array}
    \label{E6}
\end{equation}


\begin{figure}
\begin{center}
\includegraphics[width=\linewidth,keepaspectratio]{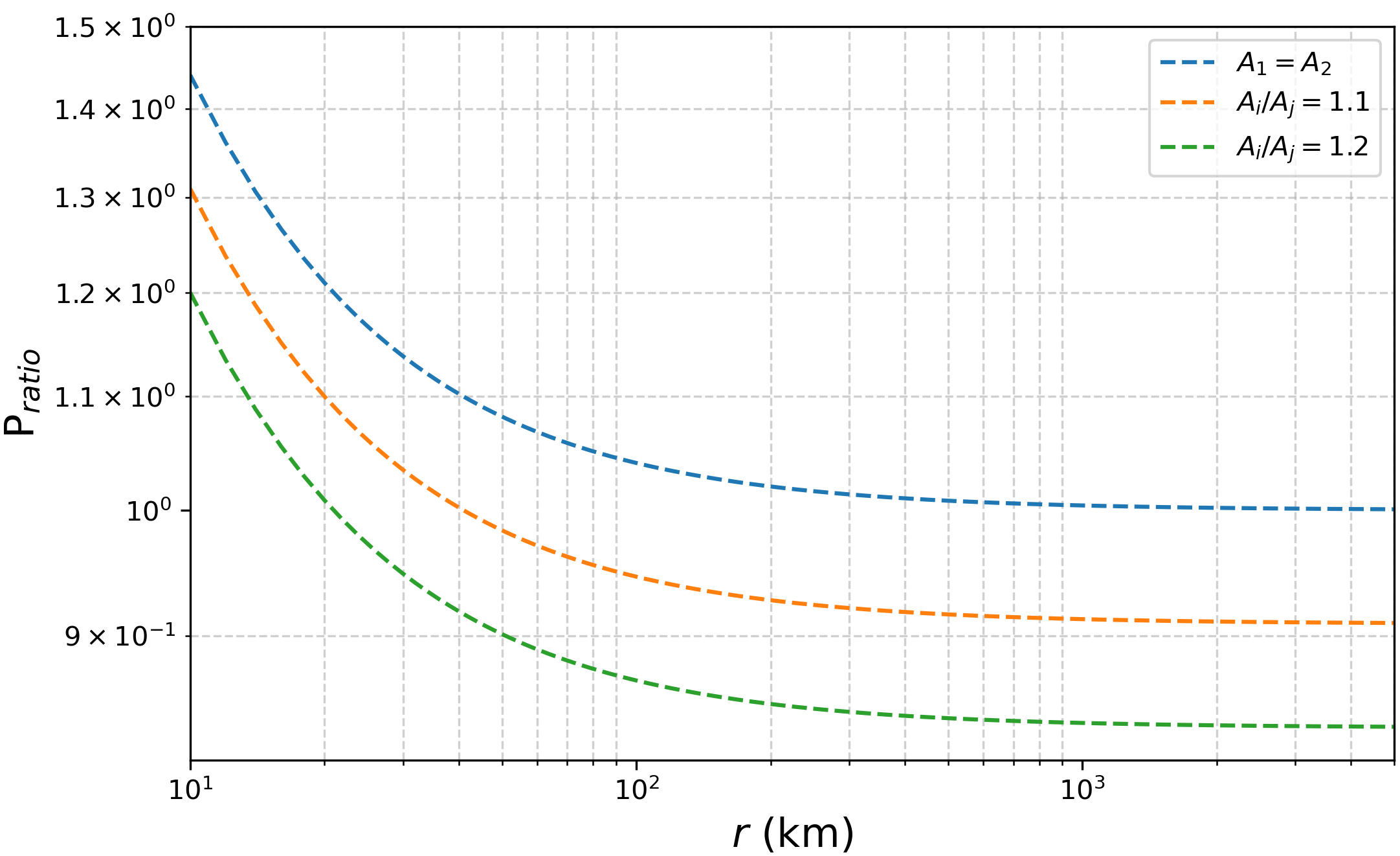}
\caption{\textbf{updated figure}We show the ratio of powers seen by a baseline with ($A_{1}$ not equal to $A_{2}$) and without ($A_{1}=A_{2}$) holographic effects. For the above plot, we assume $\Delta r = 2$\,km (approx. delay in a $6$\,km baseline observing a source $20$ degrees from the zenith). }
\label{Figamplitudefading}
\end{center}
\end{figure}

For astronomical sources, $\lim_{r \to \infty} P_{ratio} \approx \frac{A(l_{i},m_{m})}{A(l_{j}, m_{j})}$, and the two antennas only see holographic effects, but for the objects in the near-field (where $\Delta r$ is comparable to $r$) the two antennas see significantly different powers from the radiating source.   We do not account for this effect in our ISS analysis and find it to be about $P_{ratio} = 0.986$ (r=$500$\, km, $\Delta r$=$3.35$\,km) for the longer baselines.



\subsection{Confusion from near-field sources}
\label{sec:confusion}
In most radio-astronomy observations, objects in the near-field constitute Radio Frequency Interference (RFI). A common practice to mitigate RFI is to flag the data, channels and/or times, and/or baselines, that respond to the RFI signal \citep{ford2014rfi}. However, with the advent of satellite mega-constellations, the radio sky is getting increasingly polluted and we may reach a time in the future when flagging is no longer an affordable option due to the constant presence of satellite signals in the data. An alternative RFI mitigation strategy would be to subtract or `peel' the RFI's signal contribution \citep{perley2003removing} from the visibility matrix, demonstrated recently by \cite{2023arXiv230104188F} using a Bayesian framework to subtract satellite RFI from MeerKAT \citep{jonas2016meerkat} simulated data. However, based on the near-field aperture synthesis understanding developed in this work, we comment that the peeling of near-field objects without any near-field corrections could leave behind residual sidelobe confusion noise in the images, explained further below. \\

\begin{figure}
\begin{center}
\includegraphics[width=\linewidth,keepaspectratio]{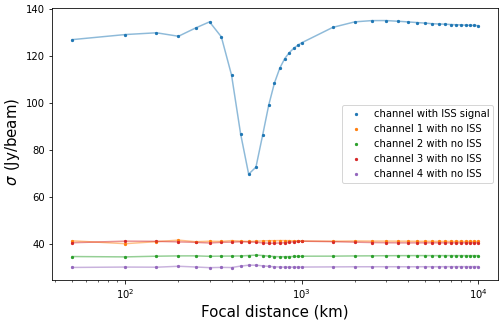}
\caption{Noise levels in residual maps for five different channels focused to a wide range of focal distances. Note that the noise is lowest in the channel with the ISS when the focal distance is approx. $500$\,km (also the line of sight range to ISS)}
\label{Figconfusion}
\end{center}
\end{figure}

We select five different fine channels (one with a strong ISS reflection signal and the others without any noticeable ISS signal), focus the visibilities to a wide range of focal distances, and plot the corresponding residual noise maps in Figure \ref{Figconfusion}. In a perfectly cleaned image, the residual map should just contain contributions from thermal noise (due to the instrument operating at non-absolute zero temperature) with a random distribution of phases, much like the bottom panels of Figure \ref{FigNullTest}, and the noise level would not be expected to change with any phase-rotation that may be applied to the visibilities. We see from Figure \ref{Figconfusion} that for the channel with the ISS signal the noise in the residual map is the lowest when the fringes are rotated to the correct focal distance of the near-field RFI, and for all other fringe positions (or focal distances) the cleaning process leaves behind residual noise due to imperfect subtraction of the near-field source. In the other four channels with no ISS signal, no astronomical sources are detected in $0.25$\,s time-averaged fine-channel data (due to very large thermal noise), and the residual map noise levels do not noticeably change with focal distance, thus demonstrating the importance of near-field considerations while subtracting near-field RFI objects from the visibility matrix. To properly reach the lowest noise levels in the data when mitigating RFI, the peeling process has to be performed in the near field. We also note that the lowest noise in the channel with the ISS signal is higher than the other channels, possibly due to increased system noise in this channel (as the ISS signal is a few thousand Jy).\\

\subsection{Fringe washing}
\label{sec:rangingprecision}


While longer baselines have better range resolution using our method (as they see more change in wavefront curvature), as previously discussed in \cite{2022AdSpR..70..812P}, the signals from fast-moving objects get blurred into smears (fringe-washed) when they move more than a synthesized beam during the visibility integration time. The phase measurement of a source by a baseline depends on the source's position in the fringe pattern projected by the baseline in the sky plane. However, for fast-moving objects such as satellites, the source moves through many fringes within the time-averaging interval of the correlator, and hence we obtain an averaged phase measurement. Hence, even though we have developed techniques in this work to account for the near-field curvature perceived by the long baselines, the reduction in SNR due to fringe-washing is not recoverable. For example, for a satellite at $400$\,km moving with an angular speed of $1^{\circ}/s$ near the zenith, apprx. $75\%$ of the MWA extended configuration baselines are fringe-washed, while even the longest baseline of MWA's compact configuration would not be affected. For a more detailed discussion on fringe-washing, we direct the reader to \cite{2022AdSpR..70..812P}.\\

\subsection{Future Work}
Near-field tools and techniques can be used in a wide range of science cases. We currently have plans to use {\tt LEOLens} to search for low-frequency intrinsic radio emission from meteors, previously only detected by \cite{obenberger2014limits} using the Long Wavelength Array \citep[][LWA]{2012JAI.....150004T}. \cite{Zhang2018LimitsMWA} previously attempted to detect the meteor emission using a $322$ hour MWA observation campaign, but no candidates were identified. The study discarded all the longer MWA baselines and just used the short baselines to mitigate the near-field effect and hence was performed at a much-reduced sensitivity. Given that we now have tools to achieve coherence on near-field objects using the long baselines, we aim to perform a more sensitive search for intrinsic meteor emission with the MWA. \\

A natural extension of the discussion on residual confusion noise from near-field objects in Section \ref{sec:confusion} is to develop a near-field RFI peeling capability in conjunction with {\tt LEOLens}. Our preliminary peeling test with the ISS observation used here shows promising results and we aim to develop this peeling method into a tool for the wider community in the future. \\

\section{Conclusions}
In this paper, we explore the near-field aperture synthesis techniques using a single observation of the ISS using the Murchison Widefield Array. For desired focal distances, we calculate the appropriate delays for every baseline which would bring the near-field signal into focus of the reconstructed image. We illustrate the effect of near-field corrections in the aperture plane that results in 'curving' the array through addition delays such that the incoming near-field wavefront falls coherently on the array. This delay correction in the aperture plane, translates to rotating fringes at different rates in the sky plane. As longer baselines see more of the near-field curvature, they undergo more delay corrections (or fringe rotation) when compared to the short baselines.  \\

Having developed a python tool that performs near-field corrections to the input interferometric dataset, we use it to demonstrate the inverse problem of inferring the range of the near-field event from the apparent curvature in radiation as seen by the array. We do so by trialing many focal distances to make a focused image of the near-field signal. When the assumed focal distance is equal (or approximately equal) to the true distance, the re-constructed image has the highest SNR. We demonstrate this using the ISS observation, and obtain ranges are in agreement with the distances predicted by its Two Line Elements. \\

We conclude the paper by discussing the limitations of the near-field methods used in this work. For objects that are very close (distances comparable to its longest baseline length) to the array, the near-field signal undergoes different amounts of propagation fading resulting in different powers seen by the two antennas of the baseline. We also find that the peeling of near-field RFI from interferometric data can leave behind residual confusion noise when not accounting for the near-field effects. 



\label{sec:conclusion}
\begin{acknowledgement}
This scientific work uses data obtained from Inyarrimanha Ilgari Bundara / the Murchison Radio-astronomy Observatory. We acknowledge the Wajarri Yamaji People as the Traditional Owners and native title holders of the Observatory site. Establishment of CSIRO's Murchison Radio-astronomy Observatory is an initiative of the Australian Government, with support from the Government of Western Australia and the Science and Industry Endowment Fund. Support for the operation of the MWA is provided by the Australian Government (NCRIS), under a contract to Curtin University administered by Astronomy Australia Limited. This work was supported by resources provided by the Pawsey Supercomputing Research Centre with funding from the Australian Government and the Government of Western Australia. Steve Prabu would also like to thank Adrian Sutinjo for the valuable comment on amplitude fading during the CIRA Journal Club talk on near-field aperture synthesis.\\

\begin{figure}
\begin{center}
\includegraphics[width=\linewidth,keepaspectratio]{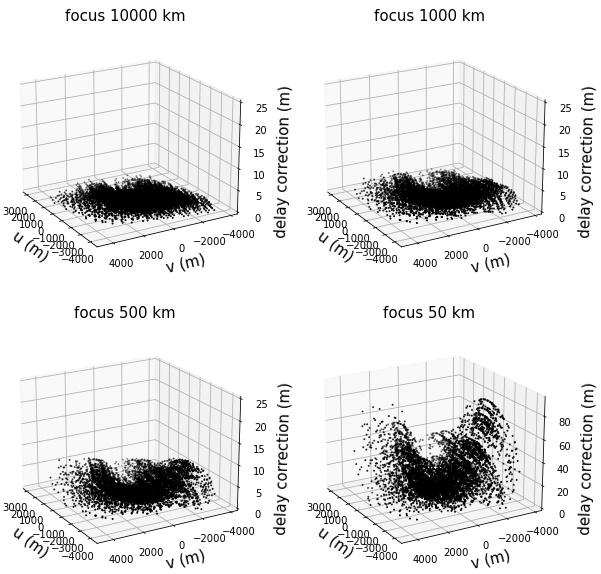}
\caption{In the above four panels we show the absolute delay correction performed by {\tt LEOLens} for four different focal distances. An animation of the figure can be obtained from \url{https://www.youtube.com/watch?v=9LIN6fErVbI}}
\label{FigUVPlane}
\end{center}
\end{figure}

\subsection*{Sofware}
We acknowledge the work and the support of the developers of the following Python packages:
Astropy \citep{theastropycollaboration_astropy_2013,astropycollaboration_astropy_2018}, Numpy \citep{vanderwalt_numpy_2011}, Scipy  \citep{jones_scipy_2001}, matplotlib \citep{Hunter:2007}, and python-casacore \footnote{\url{https://github.com/casacore/python-casacore}}. The work also used WSCLEAN \citep{offringa-wsclean-2014,offringa-wsclean-2017} for making fits images and DS9\footnote{\href{http://ds9.si.edu/site/Home.html}{ds9.si.edu/site/Home.html}} for visualization purposes. 

\end{acknowledgement}

\appendix
\section{Near-Field Correction in Aperture Plane vs Sky Plane}
\label{sec:uvwvslm}
We illustrate the effect of near-field correction in the aperture plane (u,v,w) and the sky plane (l,m) using a single FM band fine channel data for one of the time steps that the ISS was detected through reflection. \\

In the aperture plane, the near-field correction can be thought of as `curving' the array to match the near-field wavefront such that the near-field signal falls coherently on the array. We show this in the four panels of Figure \ref{FigUVPlane}. For an arbitrarily chosen four different focal distances ($10,000$\,km, $1000$\,km, $500$\,km, and $50$\,km) we show the absolute delay correction performed by {\tt LEOLens} to approx. $8000$ instantaneous baselines of the MWA. We note from Figure \ref{FigUVPlane} that for the wide range of focal distances considered, the short-baselines do not go through much delay correction as they would still see the objects in the far field. On the contrary, the long baselines have larger delay and delay rate (slope of delay vs focal distance) corrections as we bring the focal distance closer to the instrument. \\

In the sky plane, the near-field corrections result in rotating the fringes at different rates across the sky as we bring the focal distance from a faraway distance to shorter focal distances. We illustrate this using an animation provided at \url{https://www.youtube.com/watch?v=3HlUgVY_nfU}. In the top-left and top-middle panels of the animation, we show the fringe produced by a long baseline and a short baseline (the baseline lengths were arbitrarily chosen to help demonstrate better). As we bring the focal distance of the array from far-field to much closer distances, the fringes are rotated at different rates in the sky. The bottom-left and bottom-middle panel shows the delay corrections performed to the long and short baseline as we change the focal distance. The MWA has about $8000$ instantaneous baselines and in the bottom-right panel we show the combined effect of all the rotated fringes as we change the focal distance. When the assumed focal distance matches the range of the ISS (about $500$\,km), all the fringes coherently produce the image of the ISS with the maximum possible SNR.

\printendnotes

\printbibliography

\end{document}